\documentclass{PoS}
\usepackage{amssymb,amsmath,bm}
\usepackage[compress]{cite}
\usepackage{graphicx}
\usepackage{slashed}

\def \nnb{\nonumber}
\def \dps{\displaystyle}
\def \ts{\textstyle}
\def\ub{\bar{u}}

\newcommand{\calO}{\mathcal{O}}
\newcommand{\calH}{\mathcal{H}}

\newcommand{\as}{\alpha_s}

\newcommand{\hs}[1]{\hspace*{#1 pt}}
\newcommand{\vs}[1]{\vspace*{#1 pt}}

\newcommand{\eps}{\epsilon}
\newcommand{\ep}{\epsilon}

\newcommand{\nms}{\slashed{n}_-}
\newcommand{\nps}{\slashed{n}_+}

\title{
\vspace*{-3.2em}
\mbox{}\hfill \mbox{\small\sc SI-HEP-2019-25, QFET-2019-13, TUM-HEP-1244/19}\\
\vspace*{3.2em}
Non-leptonic B-decays at two loops in QCD Factorization}

\ShortTitle{Non-leptonic B-decays at two loops in QCD Factorization}

\author{Guido Bell\\%\thanks{}
       Naturwissenschaftlich-Technische Fakult\"at, Universit\"at Siegen, \\ Walter-Flex-Str.~3, 57068 Siegen, Germany\\
       E-mail: \email{bell@physik.uni-siegen.de}}

\author{Martin Beneke\\%\thanks{}
        Physik Department T31, James-Franck-Stra{\ss}e~1, \\ Technische Universit\"at M\"unchen, \\ D--85748 Garching, Germany\\
       E-mail: \email{mbeneke@tum.de}}

\author{\speaker{Tobias Huber}\\ %\thanks{}
        Naturwissenschaftlich-Technische Fakult\"at, Universit\"at Siegen, \\ Walter-Flex-Str.~3, 57068 Siegen, Germany\\
        E-mail: \email{huber@physik.uni-siegen.de}}

\author{Xin-Qiang~Li\\%\thanks{}
       Institute of Particle Physics and
Key Laboratory of Quark and Lepton Physics~(MOE),\\
Central China Normal University, \\ Wuhan, Hubei 430079, P.\ R.\ China \\
       E-mail: \email{xqli@mail.ccnu.edu.cn}}

\abstract{We report on the calculation of the two-loop penguin amplitudes in non-leptonic B decays in the framework of QCD factorization.
We discuss the computation of this genuine two-loop, two-scale problem and provide details on the matching from QCD onto SCET,
the evaluation of the master integrals, and the convolution of the hard scattering kernels with the light-cone distribution amplitude of the light meson.
Preliminary results on the size of the two-loop correction are given.}

\FullConference{14th International Symposium on Radiative Corrections (RADCOR2019)\\ 
9-13 September 2019\\
		Palais des Papes, Avignon, France}

\begin{document}

\section{Introduction}

Charmless non-leptonic decays of $B_{(s)}$ mesons play an important role in quantifying the amount of CP violation,
one of the most subtle phenomena in flavour physics. The structure of the decay amplitude can be generically written as
\begin{align}
{\cal A}(\bar B \to f) &= \lambda_u^{(D)} \, A_f^u + \lambda_c^{(D)} \, A_f^c = \sum_i \left[\lambda_{\rm CKM} \, \times \, C \, \times \, \langle f | {\cal O} | \bar B \rangle_{\rm QCD+QED} \right]_i \, .
\end{align}
The main elements of this formula are
i) the Wilson coefficients $C$ of tree or penguin operators,
ii) the CKM factors  $\lambda_p^{(D)} = V_{pb} \, V_{pD}^\ast$ which contain the weak phase, and
iii) the hadronic matrix elements  $\langle f | {\cal O} | \bar B \rangle$ which include the strong phases.
The interplay between the magnitudes and phases of these elements is at the heart of the rich and interesting
phenomenology of non-leptonic charmless B decays. Together with the plethora of data for numerous observables
such as branching ratios, direct CP asymmetries, and polarization fractions, this provides a fertile testing ground
for the CKM mechanism and for QCD effects in hadronic decays.

The task of obtaining accurate theoretical predictions for non-leptonic B decays is complicated by the purely hadronic
initial and final states, where QCD effects from many different scales arise. Several strategies have been developed
to solve this problem, mostly based on flavour symmetries of the light quarks (see e.g.~\cite{Zeppenfeld:1980ex}) or on factorization, like PQCD~\cite{Keum:2000ph,Lu:2000em} or QCD factorization (QCDF)~\cite{Beneke:1999br,Beneke:2000ry,Beneke:2001ev}.
The latter provides a rigorous and systematic framework to disentangle short- from long-distance physics in the heavy quark limit.
The factorization formula reads
\begin{align}
\dps \langle M_1 M_2 | Q_i | \bar{B} \rangle =& i m_B^2 \left\{\; F_+^{B \to
M_1}(0)
\int_0^1 du \; \; T_{i}^I(u) \;  f_{M_2} \, \phi_{M_2}(u) + \left(M_1 \leftrightarrow M_2\right) \right.\nnb \\
\dps +& \left. \int_0^\infty d\omega \int_0^1  dv du \; \;
T_{i}^{II}(\omega,v,u) \; \hat f_{B} \, \phi_B(\omega)  \; f_{M_1} \, \phi_{M_1}(v) \;
f_{M_2} \, \phi_{M_2}(u) \right\}\; .  \label{eq:QCDF}
\end{align}
The quantities $T_{i}^{I,II}$ are the perturbatively calculable hard scattering kernels, where $T_{i}^{I}$ includes the so-called vertex corrections and starts at ${\cal O}(1)$, whereas $T_{i}^{II}$ comprises the contributions from hard spectator scattering and starts at ${\cal O}(\alpha_s)$. 
The non-perturbative quantities are the $B \to M$ transition form factors $F_+^{B \to M}$, the decay constants $f_j$, and the distribution amplitudes $\phi_k$ of the heavy and light mesons. The QCDF formula~(\ref{eq:QCDF}) is valid to all orders in $\alpha_s$ and to leading power in $\Lambda_{\rm{QCD}}/m_b$. Moreover, the leading ${\cal O}(\alpha_s^0)$ term turns out to be real. Strong phases are thus either induced by perturbative contributions to the hard scattering kernels or by power-suppressed terms. Consequently, they are parametrically of order ${\cal O}(\alpha_s(m_b))$ or ${\cal O}(\Lambda_{\rm{QCD}}/m_b)$.

The matrix elements in~(\ref{eq:QCDF}) are furthermore classified in terms of topological amplitudes, on which the various
decay channels depend in different manners. For instance, one finds~\cite{Beneke:2003zv}
\begin{align}
\sqrt{2} \; \langle \pi^- \pi^0 | \, \mathcal{H}_{eff} \, | B^- \rangle \;
   & = \; A_{\pi\pi} \; \lambda_u^{(d)}   \big[\alpha_1(\pi\pi) + \alpha_2(\pi\pi) \big] \, ,\nnb \\
 - \; \langle \pi^0 \pi^0 | \, \mathcal{H}_{eff} \, | \bar{B}^0 \rangle \;
   & = \; A_{\pi\pi} \; \big\{ \lambda_u^{(d)} \big[\alpha_2(\pi\pi) - \alpha_4^u(\pi\pi)\big] - \lambda_c^{(d)}  \, \alpha_4^c(\pi\pi) \big\} \, ,\nnb \\
  \langle \pi^-\bar K^0 | \, \mathcal{H}_{eff} \, | B^- \rangle \;
   &\dps= \; A_{\pi\bar K}\, \left[\lambda^{(s)}_u \, \alpha_4^u(\pi \bar K) + \lambda^{(s)}_c \, \alpha_4^c(\pi \bar K) \right]\, .
\end{align}
Here, $\alpha_1$ and $\alpha_2$ are the colour-allowed and colour-suppressed tree amplitudes, respectively. $\alpha_4^{u,c} = a_4^{u,c} \pm r_\chi \, a_6^{u,c}$ are the QCD penguin amplitudes, with $a_4^{u,c}$ as their leading-power part. The leading order (LO) and next-to-leading order (NLO) contributions to these amplitudes have been known since a long time~\cite{Beneke:1999br,Beneke:2001ev,Beneke:2003zv} and a comprehensive phenomenological analysis based on the NLO results was carried out in~\cite{Beneke:2003zv}. At the next-to-next-to-leading order (NNLO) the one-loop ${\cal O}(\alpha_s^2)$
correction to the hard spectator scattering~\cite{Beneke:2005vv,Beneke:2006mk,Kivel:2006xc,Pilipp:2007mg,Jain:2007dy}, as well as the two-loop ${\cal O}(\alpha_s^2)$ correction to the tree topology of the vertex kernel~\cite{Bell:2007tv,Bell:2009nk,Beneke:2009ek,Bell:2009fm} have been computed more than a decade ago. The first NNLO contribution to the vertex correction of the leading QCD penguin amplitudes that became available
was the one-loop $\calO(\as^2)$ insertion of the chromomagnetic dipole operator~\cite{Kim:2011jm}.
More recently the current-current operator contribution has been completed~\cite{Bell:2015koa},
and the remaining operator insertions are currently being finalized~\cite{Bell:2020}.
With these corrections at hand one can study the impact of QCD corrections to direct CP asymmetries.

\section{Theoretical framework}

The decays of heavy quarks are described in an effective five-flavour theory where the top quark and the heavy gauge bosons $W^\pm$,~$Z$ are integrated out. The resulting effective weak Hamiltonian for $b\to D$ transitions ($D=d,s$) is given by~\cite{Buchalla:1995vs,Chetyrkin:1997gb}

\begin{align}
\calH_\text{eff} =
    +\frac{4G_F}{\sqrt{2}} \; \sum_{p=u,c} V_{pD}^* V_{pb}
    \left( C_1 Q_1^p + C_2 Q_2^p
	+ \sum_{i=3}^{6} C_i Q_i
    + C_{8g} Q_{8g} \right)
    + \text{h.c.}
\end{align}

\noindent The dimension-six operators in the so-called CMM basis~\cite{Chetyrkin:1997gb} are defined as
\begin{align}
Q_1^p &=  (\bar p_L \gamma^\mu T^A b_L ) \;  (\bar D_L \gamma_\mu T^A p_L), & Q_5   &=  (\bar D_L \gamma^\mu\gamma^\nu\gamma^\rho b_L) \; {\ts \sum_q}\; (\bar q \gamma_\mu\gamma_\nu\gamma_\rho q),\nnb\\
Q_2^p &=  (\bar p_L \gamma^\mu b_L) \; (\bar D_L \gamma_\mu p_L), & Q_6   &=  (\bar D_L \gamma^\mu\gamma^\nu\gamma^\rho T^A b_L) \; {\ts \sum_q}\; (\bar q \gamma_\mu\gamma_\nu\gamma_\rho T^A q),\nnb\\
Q_3   &=  (\bar D_L \gamma^\mu b_L) \; {\ts \sum_q}\; (\bar q \gamma_\mu q), & Q_{8g} &= \frac{-g_s}{32\pi^2} \,m_b \; \bar D \sigma_{\mu\nu} (1+\gamma_5) G^{\mu\nu} b\, .\nnb\\
Q_4   &=  (\bar D_L \gamma^\mu T^A b_L) \; {\ts \sum_q}\; (\bar q \gamma_\mu T^A q), & &
\end{align}
In dimensional regularization the operator basis needs to be supplemented by a set of evanescent operators, for which we adopt the convention of~\cite{Gambino:2003zm,Gorbahn:2004my}. The operators $Q_{1,2}^p$ are referred to as current-current operators, $Q_{3-6}$ are the QCD penguin operators, and $Q_{8g}$ is the chromomagnetic dipole operator.

The hard scattering kernels will be extracted via a matching procedure from QCD onto soft-collinear effective theory (SCET).
We denote the collinear and anti-collinear SCET fields by $\chi$ and $\xi$, respectively.
The only physical SCET operator has the fermion contraction $(\bar\chi\chi)(\bar\xi h_v)$ and is given by
\begin{align}
O_1 &= (\bar \chi \, \frac{\nms}{2} (1-\gamma_5) \chi) \; (\bar \xi \, \nps (1-\gamma_5) h_v) \, , \label{eq:scetbasis1}
\end{align}
where $h_v$ denotes the HQET heavy-quark field. In contrast, the diagrams relevant to the penguin amplitudes lead to operators where the fermion lines are contracted in the Fierz ordering $(\bar \xi \chi)(\bar\chi h_v)$. The corresponding SCET operators are chosen as
\begin{align}
\tilde O_n &=
(\bar \xi \,
\gamma_{\perp}^{\alpha}\gamma_{\perp}^{\mu_1}\gamma_{\perp}^{\mu_2}
\ldots \gamma_{\perp}^{\mu_{2 n-2}} \chi) \;
(\bar \chi (1+\gamma_5)
\gamma_{\perp\alpha}\gamma_{\perp\mu_{2 n-2}}\gamma_{\perp\mu_{2n-3}}
\ldots\gamma_{\perp\mu_1} h_v) \, .
\label{eq:scetbasis2}
\end{align}
They are evanescent for $n>1$, and $\tilde O_1$ is Fierz-equivalent to $O_1/2$ in four
dimensions. We therefore add $\tilde O_1-O_1/2$ as another evanescent SCET operator.

By imposing the matching condition one can derive master formulas for the extraction
of the hard scattering kernels $\widetilde T_i^{(\ell)}$ for each operator $Q_i$.
At tree level, one and two loops they read, respectively,
\allowdisplaybreaks{
\begin{eqnarray}
\widetilde T_i^{(0)} &=& \widetilde A^{(0)}_{i1} \, , \label{eq:mastertree}\\[0.1cm]
\widetilde T_i^{(1)} &=& \widetilde A^{(1){\rm nf}}_{i1}+ Z_{ij}^{(1)} \, 
 \widetilde A^{(0)}_{j1}
+ \underbrace{\widetilde A^{(1){\rm f}}_{i1} - A^{(1){\rm f}}_{31} \, 
 \widetilde A^{(0)}_{i1}}_{{\cal{O}}(\eps)}
- \underbrace{[\widetilde Y_{11}^{(1)}-Y_{11}^{(1)}]\, 
 \widetilde A^{(0)}_{i1}}_{{\cal{O}}(\eps)} 
- \underbrace{\sum_{b>1} \widetilde A_{ib}^{(0)} \, 
 \widetilde Y_{b1}^{(1)}}_{{\cal{O}}(\eps)} \, ,
\quad  \label{eq:master1loop} \\
\widetilde T_i^{(2)} &=& \widetilde A^{(2){\rm nf}}_{i1} + 
Z_{ij}^{(1)} \, \widetilde A^{(1)}_{j1} + Z_{ij}^{(2)} \, 
\widetilde A^{(0)}_{j1}
+ Z_{\alpha}^{(1)} \, \widetilde A^{(1){\rm nf}}_{i1}\nnb \\[0.2cm]
&& + \, (-i) \, \delta m^{(1)} \, \widetilde A^{\prime (1){\rm nf}}_{i1}
+ Z_{ext}^{(1)} \, \big[\widetilde A^{(1){\rm nf}}_{i1}+ Z_{ij}^{(1)} \, 
\widetilde A^{(0)}_{j1}\big]\nnb \\[0.2cm]
&& - \,\widetilde T_i^{(1)} \big[  C_{FF}^{(1)} + \widetilde Y_{11}^{(1)}\big] 
- \sum_{b>1} \widetilde H_{ib}^{(1)} \, \widetilde Y_{b1}^{(1)} \nnb\\[0.2cm]
&& + \,[\widetilde A^{(2){\rm f}}_{i1} - A^{(2){\rm f}}_{31} \, 
\widetilde A^{(0)}_{i1}] + \, (-i) \, \delta m^{(1)} \, 
[\widetilde A^{\prime (1){\rm f}}_{i1}
- A^{\prime (1){\rm f}}_{31} \, \widetilde A^{(0)}_{i1}]\nnb\\[0.2cm]
&& + \,(Z_{\alpha}^{(1)}+Z_{ext}^{(1)})\,
[\widetilde A^{(1){\rm f}}_{i1} - A^{(1){\rm f}}_{31} \, 
\widetilde A^{(0)}_{i1}]\nnb \\[0.2cm]
&& - \,[\widetilde M^{(2)}_{11} - M^{(2)}_{11} ] \, 
\widetilde A^{(0)}_{i1} \nnb \\[0.2cm]
&& - \,(C_{FF}^{(1)}-\xi_{45}^{(1)})\, 
[\widetilde Y_{11}^{(1)}-Y_{11}^{(1)}] \, \widetilde A^{(0)}_{i1} - 
[\widetilde Y_{11}^{(2)}-Y_{11}^{(2)}]\,
\widetilde A^{(0)}_{i1}  \nnb \\[0.2cm]
&& - \,\sum_{b>1} \widetilde A_{ib}^{(0)} \, \widetilde M_{b1}^{(2)} 
- \,\sum_{b>1} \widetilde A_{ib}^{(0)} \, \widetilde Y_{b1}^{(2)} \, . \label{eq:master2loop}
\end{eqnarray}}

The detailed derivation of the master formulas and a comprehensive explanation
of the notation can be found in~\cite{Beneke:2009ek,Bell:2020}. Here we give
merely the meaning of the most important quantities.
The $\widetilde A_{ia}^{(\ell)}$ denote bare $\ell$-loop 
on-shell matrix elements of QCD operators (index $i$), whose result
is proportional to $\tilde O_a$ on the SCET side.
For the matching procedure it turns out to be convenient to further split up 
the amplitudes $\widetilde A_{ia}^{(\ell)}$ into factorizable (superscript f) and
non-factorizable (superscript nf) diagrams, see~\cite{Beneke:2009ek}
for details.  The renormalization factors $Z_{ij}$, $Z_{\alpha}$, $\delta m$, 
and $Z_{ext}$ account for operator, coupling, mass, and wave-function
renormalization, respectively.  In the matrices $Z_{ij}$ the column index $j$
runs over both, physical and evanescent operators. 
$C_{FF}^{(1)}$~\cite{Beneke:2009ek} can be determined from one-loop matching calculations for 
heavy-to-light transitions, and the quantity $\xi_{45}^{(1)}$~\cite{Beneke:2008ei} 
gives the relation between the four- and five-flavour coupling in $D$ dimensions.
The $\widetilde Y_{ab}^{(\ell)}$ are renormalization factors in SCET which also
take the mixing of physical into evanescent SCET operators into account.

\section{The two-loop calculation}

\begin{figure}[t]
\begin{center}
\hs{8}\includegraphics[scale=.32]{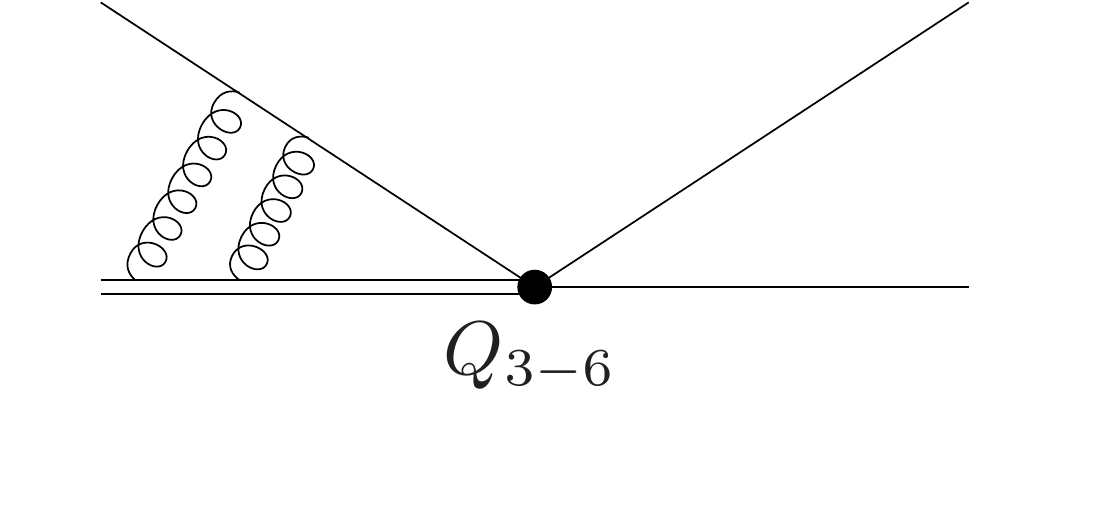}
\includegraphics[scale=.34]{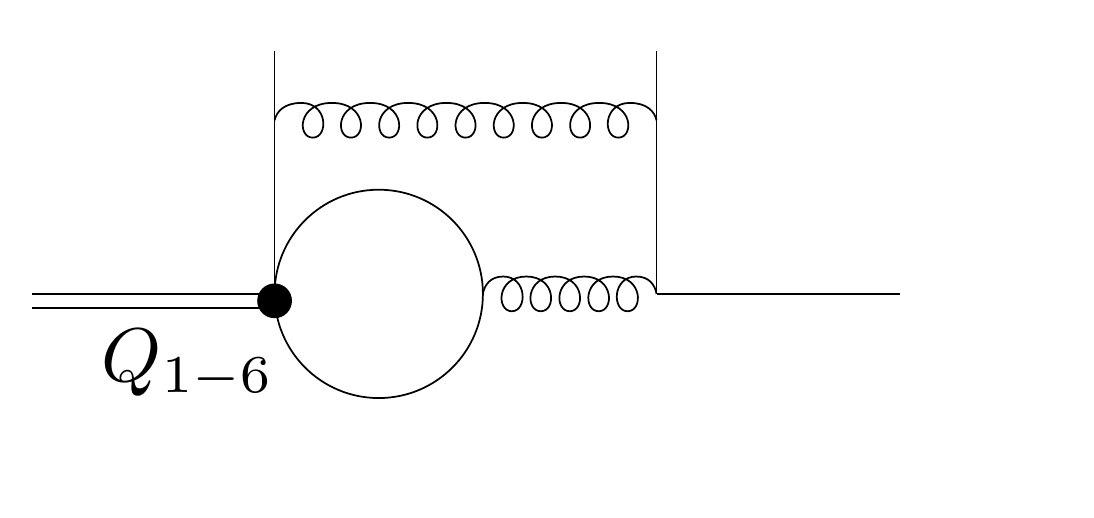}
\hs{-5}\includegraphics[scale=.34]{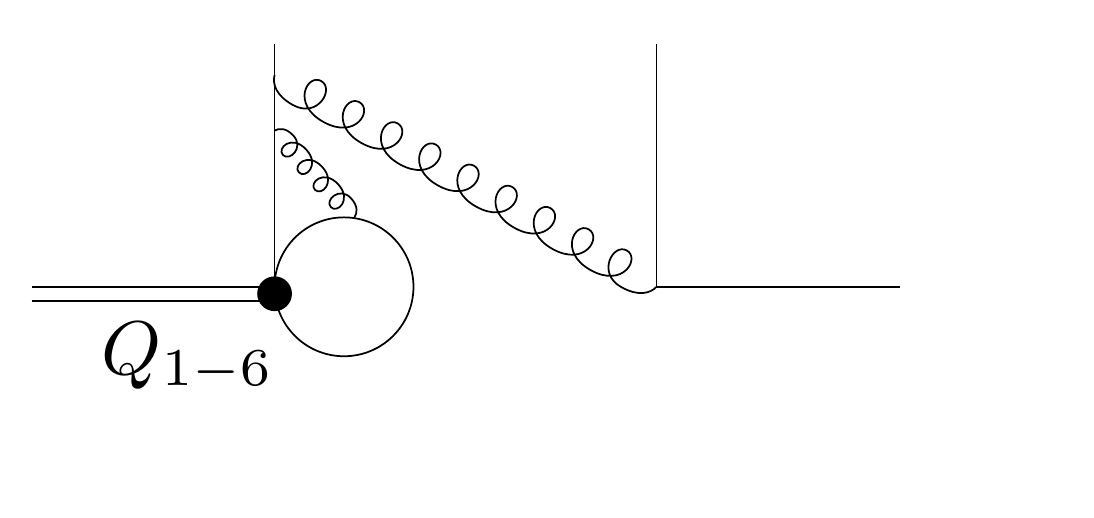}
\hs{-5}\includegraphics[scale=.34]{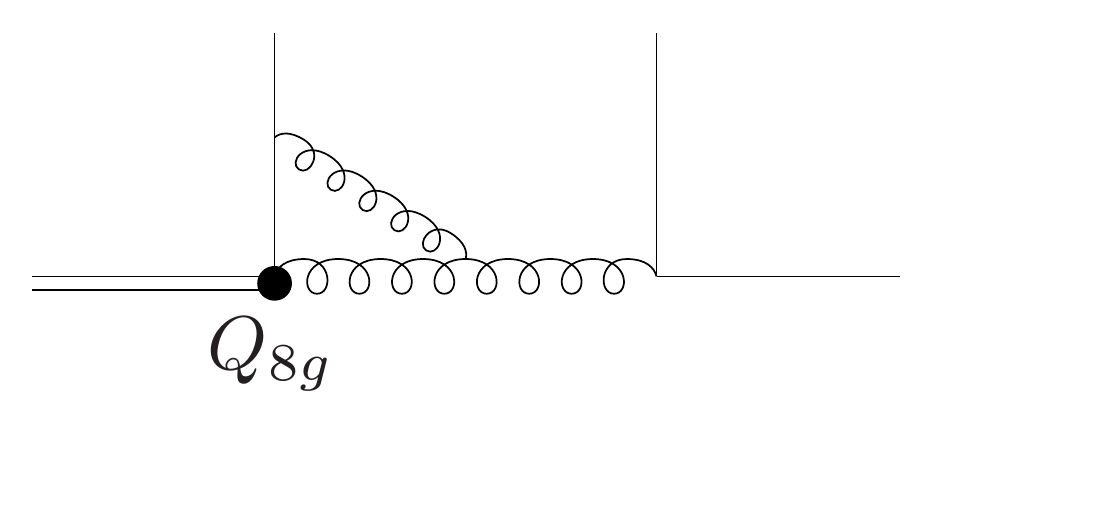}
\end{center}

\vs{-15}

\caption{\label{fig:sampleNNLO} Sample diagrams at NNLO. The diagram on the right denotes a one-loop ${\cal O}(\alpha_s^2)$ contribution from the chromomagnetic dipole operator $Q_{8g}$~\cite{Kim:2011jm}.}
\end{figure}

\begin{figure}[b]
\begin{center}
 \includegraphics[width=.99\textwidth]{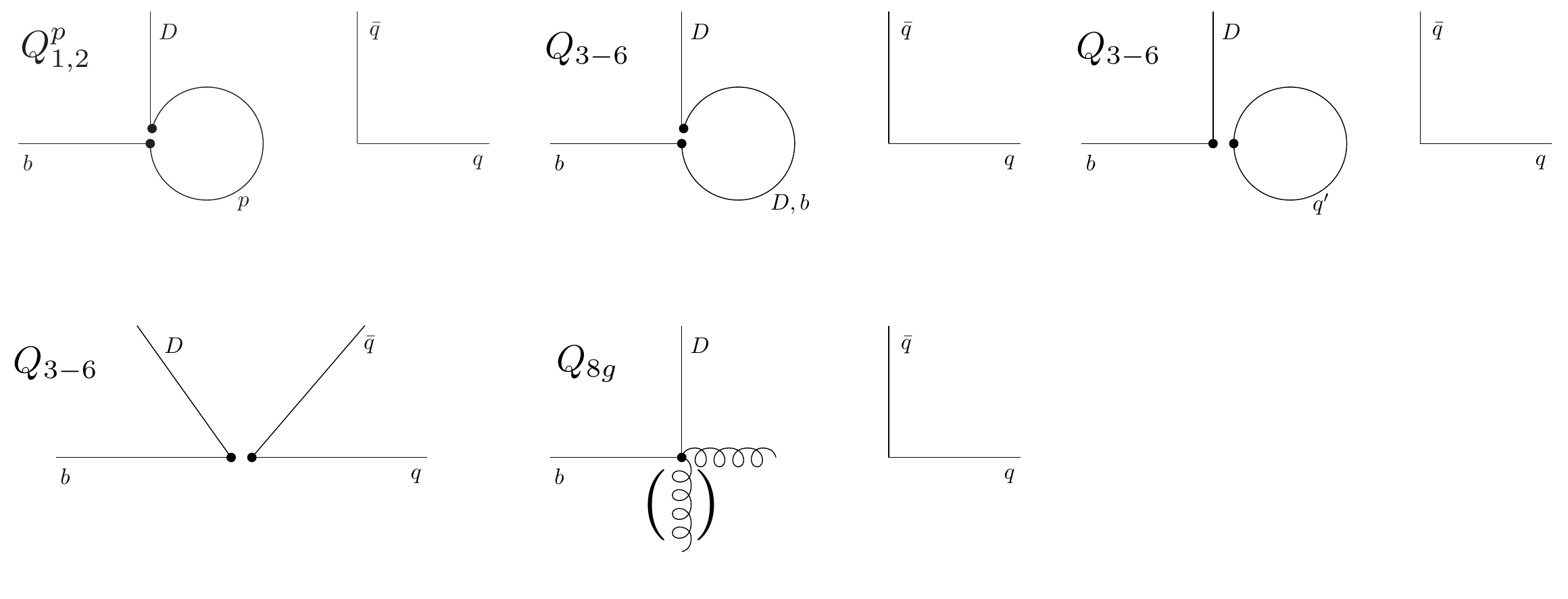}
\end{center}

\vs{-20}

\caption{\label{fig:insertions} Graphical illustration of the various operator insertions. The symbols stand for
$p \in \{u,c\}$, $D \in \{d,s\}$, $q \in \{u,d,s\}$ and $q^\prime \in \{u,d,s,c,b\}$. The black dots denote the
operator insertion from the effective weak Hamiltonian. Disconnected parts are understood to be connected by gluons.}
\end{figure}

The calculation of the NNLO correction to the vertex-kernel of the leading QCD penguin amplitudes $a_4^u$ and $a_4^c$ amounts to the evaluation of $\sim 130$ Feynman diagrams. A subset of them is shown in Fig.~\ref{fig:sampleNNLO}. The one-loop ${\cal O}(\alpha_s^2)$ contribution of the chromomagnetic dipole operator $Q_{8g}$, depicted in the right panel in Fig.~\ref{fig:sampleNNLO}, was calculated in~\cite{Kim:2011jm}. All other contributions are genuine two-loop diagrams. Moreover, the operators from the effective weak Hamiltonian contribute with several insertions to the penguin amplitudes. They are depicted in Fig.~\ref{fig:insertions} and lead to quite some bookkeeping during the calculation.

\begin{figure}[t]
\begin{center}
 \includegraphics[width=0.3\textwidth]{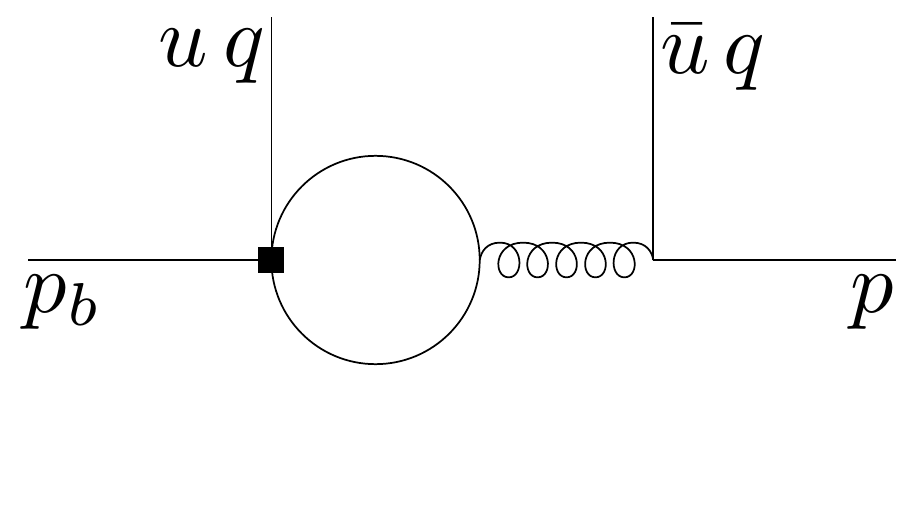}
\end{center}

\vs{-30}

\caption{\label{fig:kinematics} Kinematics of charmless non-leptonic two-body decays in QCD factorization.}
\end{figure}

The kinematics of the process can be inferred from Fig.~\ref{fig:kinematics}. The external states satisfy $p_b^2=m_b^2$ and $p^2=q^2=0$,
and the fermion in the loop (solid circle) can have mass $m_f=0$ (light quarks), $m_f=m_c$ or $m_f=m_b$.
The problem therefore depends on two dimensionless variables, the momentum fraction $\ub=1-u \in [0,1]$, and the mass ratio $z_f \equiv m_f^2/m_b^2$. There is a kinematic threshold at $\ub = 4 z_f$. For later convenience we also introduce the
variables $r =\sqrt{1-4 z_c}$ and $ s =\sqrt{1-4 z_c/\ub}$.
The reduction of the amplitude is done by techniques that have become standard in multi-loop computations. We work in dimensional regularization with $D=4-2\ep$, reduce the tensor structure via Passarino-Veltman relations, followed by IBP reduction~\cite{Tkachov:1981wb,Chetyrkin:1981qh,Laporta:2001dd} of the scalar integrals to master integrals using FIRE~\cite{Smirnov:2008iw} and an in-house routine. This procedure results in about three dozens of yet unknown master integrals.

The master integrals that stem from the insertion of penguin operators into the
``tree-type'' diagrams (left panel of Fig~\ref{fig:sampleNNLO}) were computed
more than a decade ago~\cite{Bell:2007tv,Bell:2009nk,Huber:2009se,Beneke:2009ek} and evaluate
to harmonic polylogarithms (HPLs). The remaining ones were computed analytically
in~\cite{Bell:2014zya} in a canonical basis~\cite{Henn:2013pwa} in terms of iterated integrals 
(Goncharov polylogarithms) over the alphabet
\begin{equation}
\Big\{ 0 \; , \; \pm 1 \; , \; \pm 3 \; , \; \pm i\sqrt{3} \; , \; \pm r  \; , \;  \pm\frac{r^2+1}{2} \; , \; \pm (1+2\textstyle\sqrt{z_c}) \; , \;  \pm (1-2\textstyle\sqrt{z_c}) \Big\} \, . \label{eq:alphabet}
\end{equation}

Most of the individual terms in the master formulas~\eqref{eq:mastertree}~--~\eqref{eq:master2loop}
have poles in $\ep$, which have to cancel in the total expression for the
hard scattering kernels $\widetilde T_i^{(\ell)}$. We checked the pole cancellation
analytically through to $\calO(\as^2)$ and for all operators
$i \in \{Q_1^u,Q_2^u,Q_1^c,Q_2^c,Q_{3-6},Q_{8g}\}$. Also the finite parts
of the hard scattering kernels were obtained completely analytically.

Having the hard scattering kernels completely analytically catalyses the convolution with the light-meson light-cone distribution amplitude $\phi_{M}(u)$
considerably. The latter is expanded according to
\begin{eqnarray}
\phi_{M}(u) &=& 6 \, u \, \ub \, \left[1+\sum_{n=1}^{\infty}a_n^{M} \, 
C^{(3/2)}_n(2u-1)\right] \, ,
\label{gegenbauerexp}
\end{eqnarray}
where $a_n^{M} \equiv a_n^{M}(\mu)$ and $C^{(3/2)}_n(x)$ are the Gegenbauer moments and 
polynomials, respectively. Truncating the Gegenbauer expansion at $n=2$ is sufficient in practice. 

The convolution of those parts of the hard scattering kernels that come from
``tree-type'' diagrams is known from the NNLO calculation of the tree
amplitudes~\cite{Bell:2007tv,Bell:2009nk,Beneke:2009ek} and can be done
completely analytically. For the ``penguin-type'' diagrams the majority of the terms
can be convoluted analytically by substituting $s=\sqrt{1-4 z_c/\ub}$,
\begin{align}
\int\limits_0^1 \!\! du \; \widetilde T_{i}(u) \; \phi_{M}(u) =
\int\limits_r^{+i\infty} \!\! ds \; \frac{2 s (r^2-1)}{(1-s^2)^2} \, \widetilde T_{i}(u(s)) \; \phi_{M}(u(s)) \, ,
\end{align}
because it leads to iterated integrals over the same alphabet~(\ref{eq:alphabet}).
The threshold at $\ub=4z_c$ is mapped to $s=0$. Subtleties arise when taking the limits $s \to r$
and $s \to +i\infty$. In case of the lower limit individual terms
contain power divergences proportional to $1/(r-s)^n$ which can be isolated via a Taylor expansion
about $s=r$ of the corresponding numerators and disappear in the sum of all terms.
In the case of the upper limit logarithmic divergences contained in
Goncharov polylogarithms have to be isolated, which is done via an argument inversion
with $z=i$ as pivot point. The divergences that arise in individual terms as $s \to +i\infty$
have to cancel in the end as well. This procedure yields Goncharov polylogarithms up to weight five,
which are evaluated numerically with GiNaC~\cite{Bauer:2000cp,Vollinga:2004sn}.

In this way, we obtain all $\mu$-dependent terms in the penguin amplitudes completely analytically.
The $\mu$-independent pieces are obtained as accurate interpolations in $z_c$.

\section{Preliminary results and conclusion}

\begin{figure}[t]
\begin{center}
 \includegraphics[width=.60\textwidth]{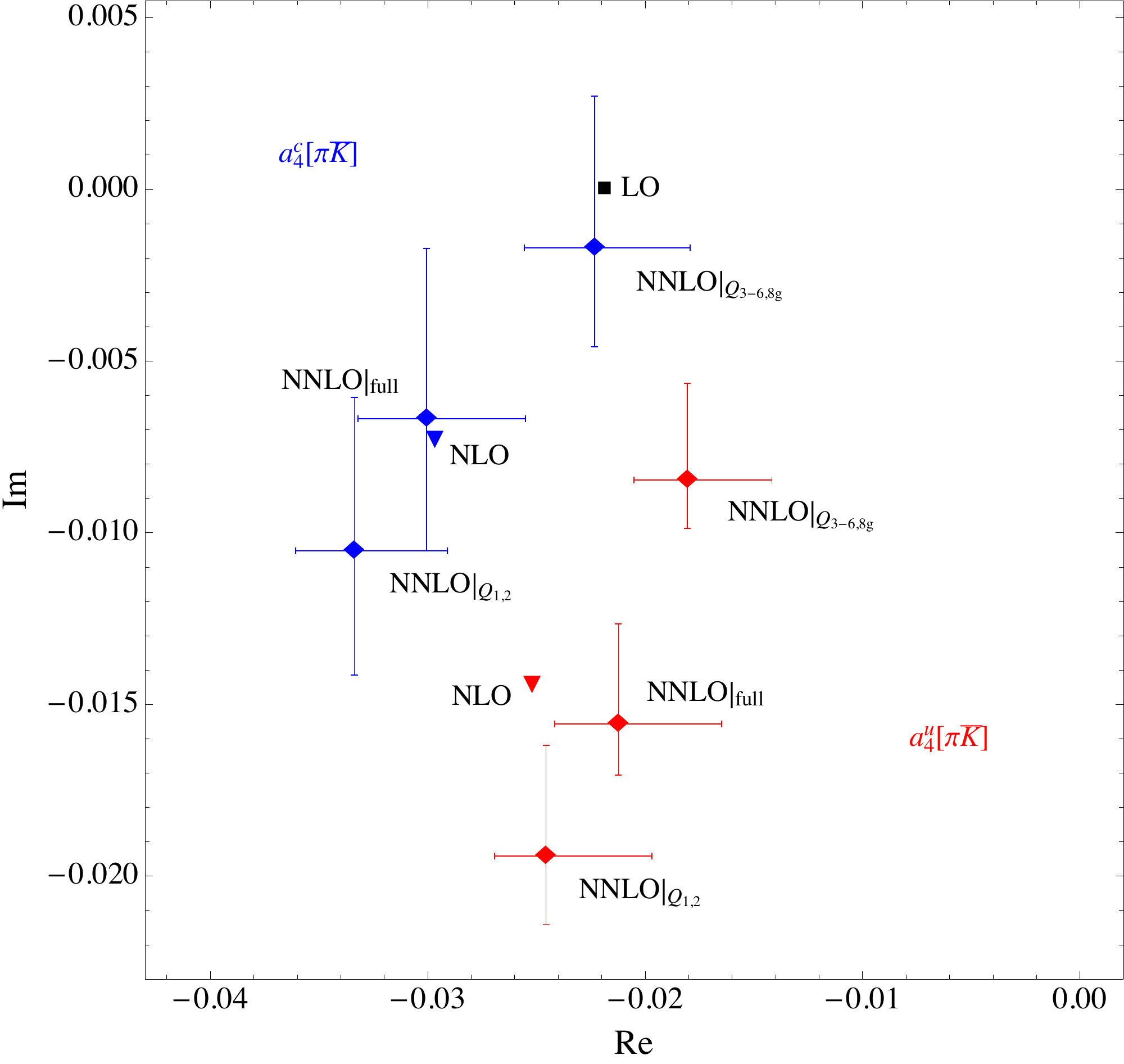}
\end{center}

\vs{-15}

\caption{\label{fig:anatomya4ua4c} Anatomy of QCD corrections to $a_4^u$ and $a_4^c$.
See text for details on the data points.}
\end{figure}

Since the publication of the results presented here is still pending, all results shown
are preliminary.
The numerical values for the leading QCD penguin amplitudes $a_4^u$ and $a_4^c$ are obtained to
NNLO by using the same input parameters as in~\cite{Bell:2015koa}, and for the $\pi \bar{K}$ final state
result in

\allowdisplaybreaks{
\begin{eqnarray}
a_4^u(\pi \bar{K})/10^{-2} &=& -2.87 -
[0.09 + 0.09i]_{\rm V_1} + [0.49-1.32i]_{\rm P_1} \nnb \\
&&- [0.32+0.71i]_{\rm P_{2}, \, Q_{1,2}^u}
+ [0.33 +0.38 i]_{\rm P_{2}, \, Q_{3-6,8g}}
\nonumber \\[0.2cm]
&& +\,\left[ \frac{r_{\rm sp}}{0.434} \right]
  \Big\{ [0.13]_{\rm LO} + [0.14 +0.12i]_{\rm HV} - [0.01-0.05i]_{\rm HP}
  + [0.07]_{\rm tw3} \Big\} \nonumber \\[0.2cm]
&=& (-2.12^{+0.48}_{-0.29}) + (- 1.56^{+0.29}_{-0.15})i\,,
\label{eq:a4unum} \\[2.0em]
a_4^c(\pi \bar{K})/10^{-2} &=& -2.87 -
[0.09 + 0.09i]_{\rm V_1} + [0.05-0.62i]_{\rm P_1} \nnb \\
&&- [0.77+0.50i]_{\rm P_{2}, \, Q_{1,2}^c}
+ [0.33 +0.38 i]_{\rm P_{2}, \, Q_{3-6,8g}}
\nonumber \\[0.2cm]
&& +\,\left[ \frac{r_{\rm sp}}{0.434} \right]
  \Big\{ [0.13]_{\rm LO} + [0.14 +0.12i]_{\rm HV} + [0.01+0.03i]_{\rm HP}
  + [0.07]_{\rm tw3} \Big\} \nonumber \\[0.2cm]
&=& (-3.00^{+0.45}_{-0.32}) + (-0.67^{+0.50}_{-0.39})i\,.
\label{eq:a4cnum}
\end{eqnarray}}

Both amplitudes receive identical contributions at leading order (LO), which is real.
The second and third term in the first line make up for the NLO result.
The second line of each equation is the NNLO result, which is split up into the
contibution from the current-current operators $Q_{1,2}^p$~\cite{Bell:2015koa} and the
penguin plus chromomagnetic dipole operators $Q_{3-6,8g}$~\cite{Bell:2020}. All terms
that multiply
\begin{align}
r_{\rm sp} =&  \frac{9 f_\pi \hat f_B}{m_b \lambda_B F_+^{B\to\pi}(0)}
\end{align}
arise from hard spectator scattering and are sensitive to the first inverse moment $\lambda_B$ of the B meson distribution amplitude.
The total error in the last line contains parametric and perturbative uncertainties added in quadrature~\cite{Bell:2015koa}.
From the NNLO numbers one can clearly see that the contribution from the penguin operators and $Q_{8g}$ tends to cancel that of the
current-current operators and therefore the NNLO result is close to the NLO one. The situation is depicted graphically in Fig.~\ref{fig:anatomya4ua4c}
where, however, the LO and NLO data points refer to the result obtained in the BBL operator basis~\cite{Buchalla:1995vs} and hence cannot be
trivially reproduced using the numbers in~(\ref{eq:a4unum}) and~(\ref{eq:a4cnum}) which were computed in the CMM basis~\cite{Chetyrkin:1997gb} .
The leading order point ``LO'' (black square) is common to both $a_4^u$ and $a_4^c$, whereas the triangles labelled ``NLO'' show the respective results through to NLO. The data points for $a_4^u$ and $a_4^c$ are indicated in red and blue, respectively.
The diamonds with error bars labelled ``NNLO'' show the results through to NNLO with only $Q_{1,2}$, only $Q_{3-6,8g}$, and all contributions included, respectively.

\begin{figure}[t]
\begin{center}
\includegraphics[width=0.99\textwidth]{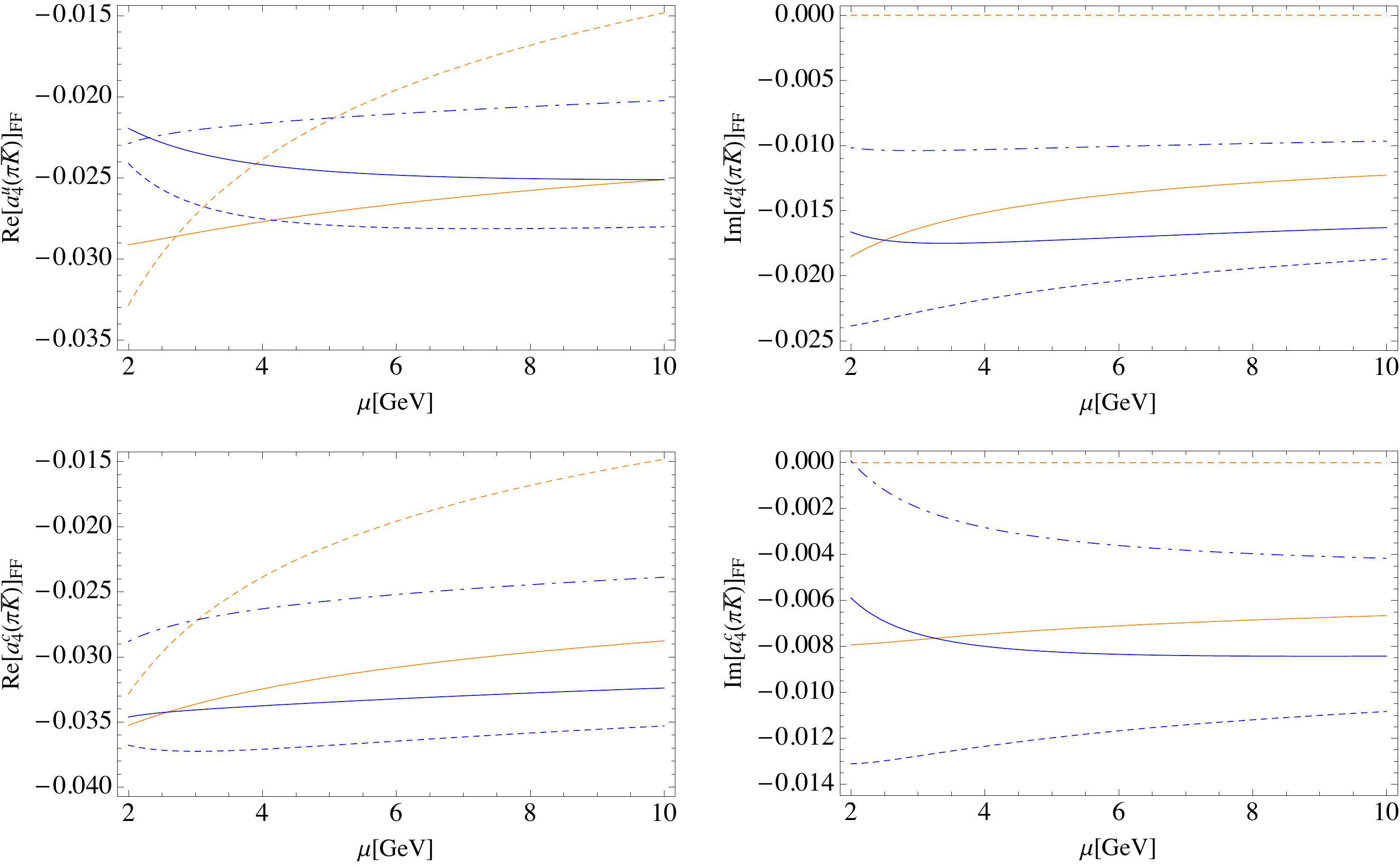}
\end{center}

\vs{-15}

\caption{\label{fig:scalea4} Scale dependence of $a_4^u$ (upper two panels) and  $a_4^c$ (lower two panels). The curves show only the
form factor (FF) contribution without spectator scattering. Orange lines represent results to LO (dashed) and NLO (solid), respectively. The dashed (dash-dotted) blue lines denote NNLO results with only current-current (only penguin and chromomagnetic dipole) operators included, while the blue solid lines show the full NNLO result.}
\end{figure}

Fig.~\ref{fig:scalea4} shows the scale dependence of the vertex correction of $a_4^u$ (upper two panels) and  $a_4^c$ (lower two panels).
Note that, like in Fig.~\ref{fig:anatomya4ua4c}, the LO and NLO curves refer to the result obtained in the BBL basis.
One can clearly see a reduction of the scale dependence from the LO (orange dashed) to the NLO (orange solid) and finally the NNLO (blue solid)
curves. The reduction is slightly less pronounced in the imaginary parts since NNLO is in fact only their first correction.

To conclude, we completed the calculation of the leading penguin amplitudes in QCD factorization to NNLO.
The computation is done via a matching from QCD onto SCET and is complicated by the presence of evanescent operators in both theories.
During the calculation one has to deal with a genuine two-loop, two-scale problem which we solved almost completely analytically.
The NNLO correction turns out to be numerically small, with a cancellation between the contributions from current-current and penguin operators.
The precise impact of the NNLO correction on the phenomenology of charmless two-body non-leptonic B decays will be investigated in the future.

\section*{Acknowledgments}

We would like to thank the organisers of ``RADCOR 2019'' for creating a very pleasant and inspiring atmosphere. The work of GB and TH was supported by DFG Forschergruppe FOR 1873 ``Quark Flavour Physics and Effective Field Theories''. The work of MB is supported by the DFG Son\-der\-for\-schungs\-bereich/Trans\-regio~110 ``Symmetries and the Emergence of Structure in QCD''.
The work of XL is supported by the National Natural Science Foundation of China under Grant Nos.~11675061 and 11435003,
and by the Fundamental Research Funds for the Central Universities under Grant No.~CCNU18TS029.

%\begin{thebibliography}{99}
%\bibitem{...}

%\bibliographystyle{JHEP}
%\bibliography{references}

\providecommand{\href}[2]{#2}\begingroup\raggedright\endgroup

%\end{thebibliography}

\end{document}